\documentclass[useAMS,usenatbib]{mn2e}

\usepackage{epsfig}
\usepackage{graphicx}
\usepackage{multirow}
\usepackage{amssymb,amsmath}
\usepackage{url}
\usepackage{float}

\title[Adaptive optics observations of the gravitationally lensed quasar SDSS~J1405+0959]{Adaptive optics observations of the gravitationally lensed quasar SDSS~J1405+0959\thanks{Based on data collected at Subaru Telescope, which is
operated by the National Astronomical Observatory of Japan.}}
\author[C.E.~Rusu et al.]
{Cristian E. Rusu,$^{1}$\thanks{E-mail: eduard.rusu@nao.ac.jp}
Masamune Oguri,$^{2,3,4}$\thanks{E-mail: masamune.oguri@ipmu.jp} 
Yosuke Minowa,$^{5,7}$ 
Masanori Iye,$^{1,6,7}$
 \newauthor
Anupreeta More$^{4}$,
Naohisa Inada$^{8}$ 
and
Shin Oya$^{5}$ \\
$^1$Optical and Infrared Astronomy Division, National Astronomical
Observatory of Japan, 2-21-1 Osawa, Mitaka, Tokyo 181-8588, \\ 
Japan\\ 
$^2$Research Center for the Early Universe, University of Tokyo,
7-3-1 Hongo, Bunkyo-ku, Tokyo 113-0033, Japan\\
$^3$Department of Physics, University of Tokyo, 7-3-1 Hongo,
Bunkyo-ku, Tokyo 113-0033, Japan\\
$^4$Kavli Institute for the Physics and Mathematics of the Universe
(Kavli IPMU, WPI), University of Tokyo, 5-1-5 Kashiwanoha,\\  
Kashiwa-shi,  Chiba 277-8583, Japan\\
$^5$Subaru Telescope, National Astronomical Observatory of Japan,
	       650 North A'ohoku Place, Hilo, Hawaii 96720, USA\\
$^6$Department of Astronomy, University of Tokyo, 
7-3-1 Hongo, Bunkyo-ku, Tokyo 113-0033, Japan\\
$^7$Department of Astronomical Science, The Graduate University
                for Advanced Studies (SOKENDAI), National Astronomical\\ 
                Observatory of Japan, 2-21-1,Osawa, Mitaka, Tokyo 181-8588, Japan\\                 
$^8$Department of Physics, Nara 
	      National College of Technology, Yamatokohriyama,
	      Nara 639-1080, Japan\\    
}

\begin{document}

%\date{Accepted 1988 December 15. Received 1988 December 14; in original form 1988 October 11}

\pagerange{\pageref{firstpage}--\pageref{lastpage}} \pubyear{2014}

\maketitle

\label{firstpage}

\begin{abstract}
We present the result of Subaru Telescope multi-band adaptive optics observations of the complex gravitationally lensed quasar SDSS~J1405+0959, which is produced by two lensing galaxies. These observations reveal dramatically enhanced morphological detail, leading to the discovery of an additional object 0\farcs26 from the secondary lensing galaxy, as well as three collinear clumps located in between the two lensing galaxies. The new object is likely to be the third quasar image, although the possibility that it is a galaxy cannot be entirely excluded. If confirmed via future observations, it would be the first three image lensed quasar produced by two galaxy lenses. In either case, we show based on gravitational lensing models and photometric redshift that the collinear clumps represent merging images of a portion of the quasar host galaxy, with a magnification factor of $\sim 15-20$, depending on the model. 
\end{abstract}


%%%%%%%%%%%%%%%%%%%%%%%%%%%%%%%%%%%%%%%%%%%%%%%%%%%%%%%%%%%%%%%%%%%%%%

\section{Introduction}

The overwhelming majority of the more than 100 gravitationally lensed quasars known so far consist of two quasar images or, due to the ellipticity of the gravitational potential in the lensing mass, four images. These are galaxy-scale lensed systems, with image separations of $\sim1\arcsec-2\arcsec$, produced by a foreground galaxy acting as a lens. Rarely, small-separation three-image lensed quasars have also been observed, \citep[e.g.,][]{lewis02,winn04}, but these contain a central quasar image, close to the center of a lensing galaxy with a central density profile that is shallower than usual. Ignoring large-separation lensed quasars produced by galaxy clusters \citep[e.g.,][]{oguri08}, another way of producing other than two or four quasar images is found in systems with a group of two \citep[e.g.,][]{shin08} or three lensing galaxies, with a single confirmed example being the six-image B1359+154 \citep[][]{rusin01}. Such systems also offer the chance to study in detail the mass distribution in galaxy pairs from the strong lens modeling of the multiple quasar images, as well as the shape of extended arcs from the quasar host galaxy. 

An example of a unique lensed quasar in a two-lens system was MG 2016+112 \citep[e.g.,][]{lawrence84,koopmans02,more09}, which contains an observed arc that has no optically detected central quasar component, and requires the consideration of both lensing galaxies to explain the observed configuration. 
Here we show that the recently discovered system SDSS~J1405+0959 \citep[][]{jackson12,inada14} is the second example of such a system, and is also likely to be a three-image lensed quasar system. This result is a product of an adaptive optics (AO) imaging campaign of gravitationally lensed quasars \citep[][Rusu et al., in prep.]{rusu11} from the SDSS Quasar Lens Search \citep[SQLS;][]{oguri06,inada12}. 

In Section \ref{sect:previous} we summarize the previous observations of this object. In Section \ref{sect:AOoverall} we present our adaptive optics observations, data reduction and morphological modeling technique. Section \ref{section:photoz} addresses the new multi-band color information, and Section \ref{section:lens} presents our gravitational lens mass models. We summarize our conclusions in Section \ref{sect:concl}. Throughout this work, the concordance cosmology with $H_0=70\ \mathrm{km}^{-1}\  \mathrm{s}^{-1}\ \mathrm{Mpc}^{-1}$, $\Omega_m=0.27$ and $\Omega_\Lambda=0.73$ is assumed.

%%%%%%%%%%%%%%%%%%%%%%%%%%%%%%%%%%%%%%%%%%%%%%%%%%%%%%%%%%%%%%%%%%%%%%

\section{Previously available observations}\label{sect:previous}

SDSS~J1405+0959 was identified in the SDSS Data Release 7 \citep{abazajian09} as a two-component system, one of which is a quasar at $z=1.810$. The system was reported by \citet{jackson12} as a new gravitationally lensed quasar discovered in the Major UKIDSS--SDSS Cosmic Lens Survey (MUSCLES). This was based on the spectral similarity of the two components, as well as the decrease in separation with increasing wavelength, suggesting the presence of a lensing galaxy. Based on a tentative detection of the $4000$ \AA\ break, as well as possible emission lines, the galaxy redshift was estimated to be $z\sim0.66$. 

The system was identified independently as a lensed quasar candidate from the SQLS \citep[see][]{inada12}, with follow-up spectroscopy and $V$, $R$, $I$ band imaging confirming the gravitational lens nature \citep{inada14}. An additional close-by component, also visible in \citet{jackson12}, in the infrared, was identified as a potential secondary lens (see Figure \ref{fig:1405closeup}). 

%%%%%%%%%%%%%%%%%%%%%%%%%%%%%%%%%%%%%%%%%%%%%%%%%%%%%%%%%%%%%%%%%%%%%%

\section{Adaptive optics observations}\label{sect:AOoverall}

\subsection{Description of the observations and data reduction}\label{sect:AO}

Adaptive optics observations were performed with the Infrared Camera and Spectrograph \citep[IRCS;][]{kobayashi00} at the 
Subaru Telescope, in conjunction with the Laser Guide Star Adaptive Optics system \citep[LGS$+$AO188;][]{hayano08,hayano10,minowa12}. IRCS was used in the 0\farcs0528 pixel scale mode, providing a field of view of $54\arcsec\times54\arcsec$. AO188 employs a curvature wavefront sensor and a 188-element bimorph deformable mirror. The LGS system uses an artificial sodium laser guide star for high-order wavefront sensing \citep{watanabe04}.

Observations were initially performed in the $K'$ band, followed a year later by $J$ and $H$ band, in order to obtain photometric redshifts for the newly detected objects. A summary of the exposure times, observation dates, airmass and photometric standard stars is provided in Table \ref{tab:followup-data}.

%%%%%%%%%%%%%%%%%%%%%%%%%%%%%%%%%%%%%%%%%%%%%%%%%%%%%%%%%%%%%%%%%%%%%%

\begin{table*}
 \centering
 \begin{minipage}{113mm}
  \caption{Summary of observations}
  \begin{tabular}{@{}cclcr@{}}
  \hline 
Filter &
Exposure [s] &  
Observation date (UTC) &
Airmass &
Photometric standard \\ 
 \hline
 $K'$ & $14\times60$ & 2012 February 21 & 1.10-1.14 & FS23 \\
$J$ & $19\times60$ & 2013 April 27 & 1.02 & FS126  \\
$H$ & $26\times60$ & 2013 April 27 & 1.02-1.07 & FS126 \\
\hline
\end{tabular}
\\ 
%{\footnotesize aaa}
%\bigskip
\label{tab:followup-data}
\end{minipage}
\end{table*}

%%%%%%%%%%%%%%%%%%%%%%%%%%%%%%%%%%%%%%%%%%%%%%%%%%%%%%%%%%%%%%%%%%%%%%

Even with the use of a laser guide star, a natural bright star is still required in the proximity of the science target, in order to perform low-order, tip-tilt mode wavefront corrections. Such a tip-tilt star was identified well within the recommended brightness limit of $R=18$ mag, and inside the recommended separation range of $80\arcsec$ (Table \ref{tab:AO}). In addition, as a bright star does not exist in the SDSS~J1405+0959 field of view, a separate star located $\sim30\arcmin$ away was observed just after the system (only in the $H$ band), as an independent point-spread function (PSF) estimator. This star was selected to have a similar tip-tilt star to that of the system, in terms of brightness and separation (Table \ref{tab:AO}). 

%%%%%%%%%%%%%%%%%%%%%%%%%%%%%%%%%%%%%%%%%%%%%%%%%%%%%%%%%%%%%%%%%%%%%%%

\begin{table*}
 \centering
 \begin{minipage}{114mm}
  \caption{Tip-tilt and PSF stars}
  \begin{tabular}{@{}cllr@{}}
  \hline 
Object &
Description &
Magnitude &
Separation [\arcsec] \\ 
 \hline
SDSS~J140514.96+095820.10 & tip-tilt star & R=13.35 & $\sim$71.5\\
SDSS~J140517.09+103007.16 & tip-tilt for the PSF star & R=13.1 & 70.8\\
SDSS~J140521.31+102933.31 & PSF star & K=12.0 &\\
\hline
\end{tabular}
\\ 
{\footnotesize The magnitudes are taken from the Naval Observatory Merged Astrometric Dataset \citep[NOMAD;][]{zacharias04} catalogue. "Separation" specifies the distance between the tip-tilt star and the object, and between the tip-tilt and the PSF star, respectively.}
%\bigskip
\label{tab:AO}
\end{minipage}
\end{table*}

%%%%%%%%%%%%%%%%%%%%%%%%%%%%%%%%%%%%%%%%%%%%%%%%%%%%%%%%%%%%%%%%%%%%%%%

The observations were performed with $8\arcsec$ dithering, in order to remove bad pixels and cosmic rays, and to allow for the creation of flat frames and sky frames from the data. Data reduction was performed with IRAF\footnotemark  \footnotetext{IRAF is distributed by the National Optical Astronomy Observatory, which is operated by the Association of Universities for Research in Astronomy (AURA) under cooperative agreement with the National Science Foundation.}, using the IRCS IMGRED\footnotemark \citep{minowa05}\footnotetext{The package is available at \protect\url{http://www.naoj.org/Observing/DataReduction/index.html}} package designed to reduce data obtained with IRCS, and supplemented with a geometrical distortion correction. 

Photometric zero-point calibration was performed using the observed standard stars in Table \ref{tab:followup-data}. All objects were corrected for Galactic extinction \citep{schlegel98}, and atmospheric extinction relative to the standard star. The standard star catalogue magnitudes at $K$ band were used to calibrate the photometry at $K'$ band, since the expected differences based on interpolation\footnotemark \footnotetext{\protect\url{http://www2.keck.hawaii.edu/inst/nirc/UKIRTstds.html}} are small ($0.01-0.03$ mag). 

%%%%%%%%%%%%%%%%%%%%%%%%%%%%%%%%%%%%%%%%%%%%%%%%%%%%%%%%%%%%%%%%%%%%%%%

\subsection{PSF and morphological modeling}\label{sect:PSF}

\begin{figure}
\includegraphics[width=84mm]{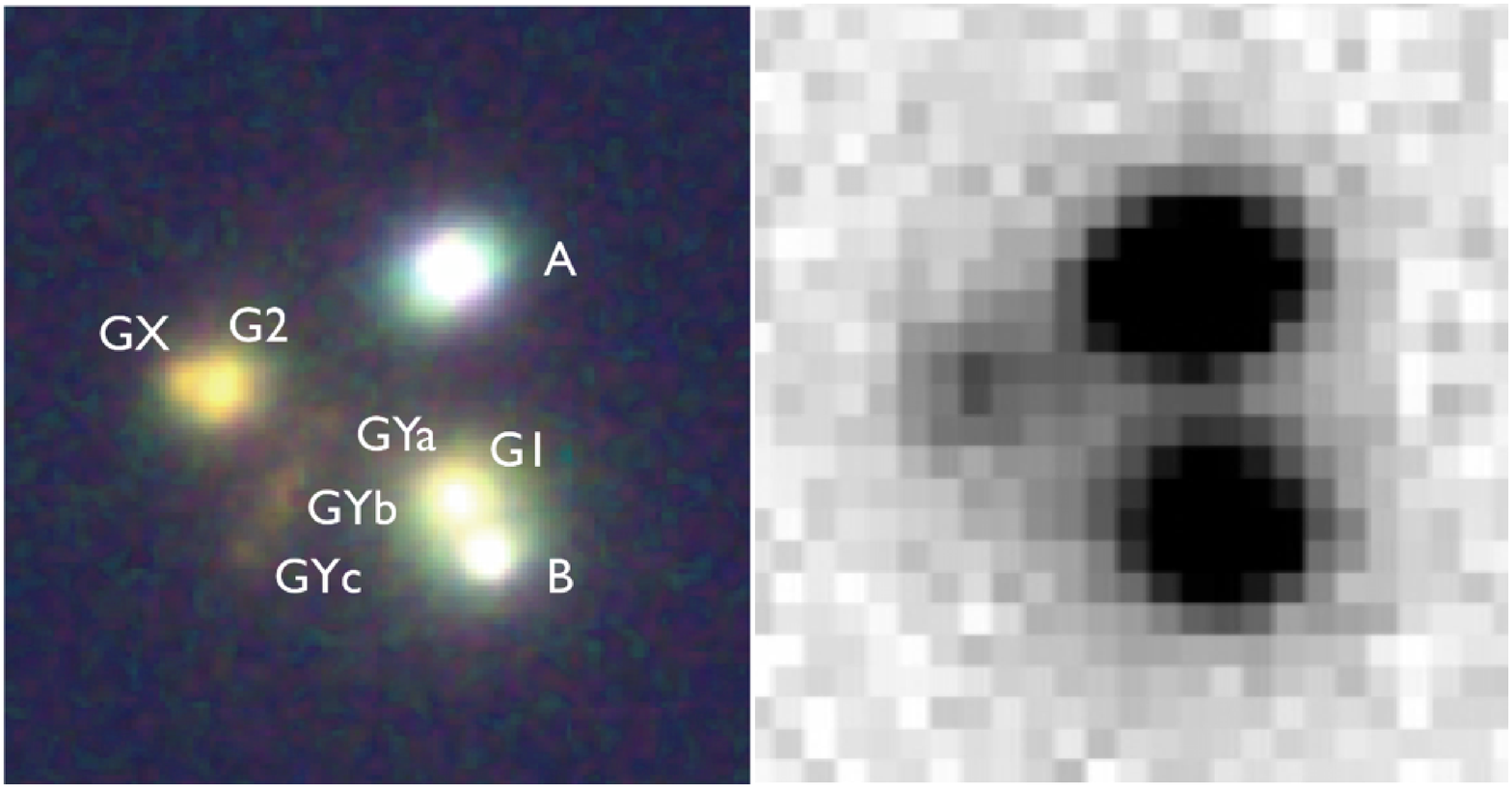}
\caption{ $\mathit{Left:}$ $5\arcsec\times5\arcsec$ $JHK'$ color combined close-up of SDSS~J1405+0959. The labels correspond to the text. $\mathit{Right:\ } $$I-$band image of the system from the discovery paper \citep[][UH88/8k, $0\farcs235$ pixel scale, 450 s]{inada14}. North is up and East is to the left.
\label{fig:1405closeup}}
\end{figure}

 As seen from Figure \ref{fig:1405closeup}, the adaptive optics observations reveal dramatically increased detail compared to \citet{inada14}. Two new features were identified. First, there is a red object marked GX just to the East of galaxy G2. Second, there is a fainter highly elongated object, GY, in between G1 and G2, which seems to be fragmented into three components, the central one being brightest. As the same structure was observed in both $H$ and $K'$ band, this fragmentation is real, and not due to low signal-to-noise (S/N).
 
 %%%%%%%%%%%%%%%%%%%%%%%%%%%%%%%%%%%%%%%%%%%%%%%%%%%%%%%%%%%%%%%%%%%%%%%
%%%%%%%%%%%%%%%%%%%%%%%%%%%%%%%%%%%%%%%%%%%%%%%%%%%%%%%%%%%%%%%%%%%%%%%

\begin{figure}
\includegraphics[width=84mm]{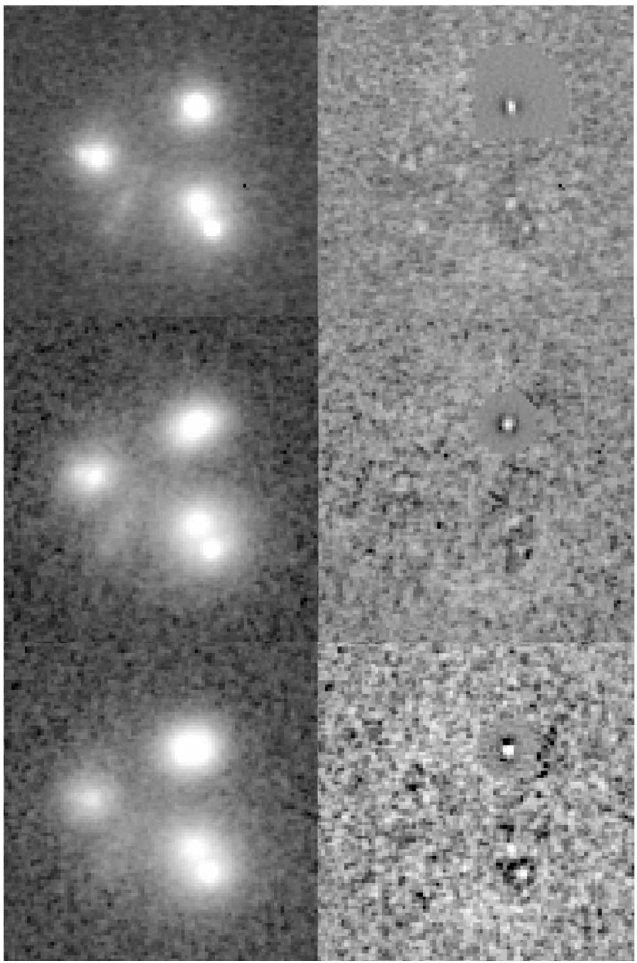}
\caption{SDSS~J1405+0959 imaging observations in the $K'$, $H$ and $J$ bands (from $\mathit{top}$ to $\mathit{bottom}$). Shown are original $5\arcsec\times5\arcsec$ frames $\mathit{(left)}$ and residuals after Galfit modeling using a hybrid PSF built on image A $\mathit{(right)}$. GY has been modeled as a single Sersic profile.
\label{fig:1405resid}}
\end{figure}

%%%%%%%%%%%%%%%%%%%%%%%%%%%%%%%%%%%%%%%%%%%%%%%%%%%%%%%%%%%%%%%%%%%%%%

 In order to characterize the morphology of the individual components of this system, as well as extract relative astrometric and photometric quantities, it is necessary to have an accurate knowledge of the PSF of the system. The PSF is difficult to characterize for most adaptive optics observations, as it varies significantly on scales of $\lesssim 1\arcmin$, or the size of the isoplanatic angle. On these scales, the probability of serendipitously finding a bright star on the line of sight to extragalactic sources is very small. In the present case, a single star is found reasonably close to the system, but is significantly fainter than the point-like quasar components. Therefore, as described in Section \ref{sect:AO}, we attempted to characterize the PSF by observing a separate bright star. However, as was noticed in other works \citep[e.g.,][]{kuhlbrodt05} which have employed this technique, temporal changes in the atmospheric turbulence characteristics induce different response from the AO system, making the use of non-simultaneous PSF estimates problematic. Indeed, using the star as a PSF template has resulted in outstanding residuals when modeling away the point-like components in SDSS~J1405+0959. For this reason, we have skipped the separate PSF star observations in the $J$ and $K'$ bands altogether. Nonetheless, we have still made use of the PSF star observations, as will be shown below. 
 
Since a suitable external PSF estimate is not available, the information provided by the observed system itself must be used in order to estimate the PSF \citep[e.g.,][]{lagattuta10,koptelova13,oya13}. We adopted a new technique, in which for each of the three-band observations, we started by masking G2, GX and GY, then we fitted the bright components A, B and G1 simultaneously. Here, A and B consisted of an analytical PSF approximated as two concentric elliptical Moffat profiles, and G1 was the convolution between the analytical PSF and a Sersic profile. The best-fit analytical parameters were obtained through $\chi^2$ minimization, based on the method implemented in {\it glafic} \citep{oguri10}. The resulting first-order PSF was used to fit G2, GX and GY as well, each as a Sersic profile, and subtract them. After masking the prominent residuals left under the subtracted objects, the remaining component A was refitted with an analytical PSF. The best-fit PSF parameter values are given in Table \ref{tab:analpsf}. A final hybrid PSF, consisting of a core and a wing component, was then created by keeping the observed distribution of pixel values in a circular region centered around the peak, large enough to include all visible non-analytical residuals, and replacing the distribution at larger radii with the analytical profile. This technique produces a PSF that is cleared of contamination from the nearby objects, includes the non-analytical features that are otherwise not reproduced by an analytical PSF, and has a wing that is noise-free. Components A, B, G1, G2, GX and GY were then refitted with Galfit \citep{peng02} using the new PSF, where the quasar images A and B were fitted as PSFs, and G1, G2 as Sersic profiles, convolved with the PSF. GX was fitted as a PSF and, alternatively, a Sersic profile, whereas GY was fitted either as a single Sersic profile, or as a Sersic profile (GYb) and two PSFs (GYa and GYc). The residuals after morphological modeling in the three bands are shown in Figure \ref{fig:1405resid}. The extracted astrometry and photometry, as well as the parameters of the extended components, are given in Tables \ref{tab:1405astrometry}, \ref{tab:1405photometry}  and \ref{tab:1405morphology}.

 %%%%%%%%%%%%%%%%%%%%%%%%%%%%%%%%%%%%%%%%%%%%%%%%%%%%%%%%%%%%%%%%%%%%%%
 
\begin{table*}
 \centering
 \begin{minipage}{143mm}
  \caption{Analytical parameters of the two-component Moffat PSFs}
  \begin{tabular}{@{}lcccccccccc@{}}
  \hline 
 \hline
\multirow{2}{*}{Band} &
\multirow{2}{*}{{\it FWHM1}} &
\multirow{2}{*}{{\it e1}} &
\multirow{2}{*}{{\it PA1}} &
\multirow{2}{*}{{\it $\beta 1$}} &
\multirow{2}{*}{{\it FWHM2}} &
\multirow{2}{*}{{\it e2}} &
\multirow{2}{*}{{\it PA2}} &
\multirow{2}{*}{{\it $\beta 2$}} &
flux1/ &
Strehl \\ 
& & & & & & & & & (flux1 + flux2) & ratio ($\%$) \\
\hline
$J$ & $0.18$ & $0.07$ & $21.7$ & $6.0$ & $0.55$ & 0.16 & $-75.9$ & 2.5 & 0.15 & 2 \\
$H$ & $0.17$ & $0.07$ & $-48.0$ & $8.3$ & $0.51$ & 0.27 & $-63.2$ & 2.2 & 0.16 & 4 \\
$H$ (PSF) & $0.17$ & $0.09$ & $12.8$ & $43.3$ & $0.46$ & 0.14 & $-42.5$ & 2.3 & 0.27 & 6 \\
$K'$ & $0.18$ & $0.11$ & $27.9$ & $6.3$ & $0.54$ & 0.07 & $-60.4$ & 1.7 & 0.21 & 7 \\
\hline
\end{tabular}
\\ 
{\footnotesize Affix $\mathit{1}$ refers to the core, and affix $\mathit{2}$ refers to the wings, both modeled as Moffat profiles. Here $\mathit{FWHM}$ ($\arcsec$), $\mathit{e}$, $\mathit{PA}$ (deg) and $\beta$ are the profile full width at half maximum, ellipticity, position angle (positive from North towards East), and $\beta$ parameter, respectively. The Strehl ratios were computed by comparing the peak flux of the PSF with that of a diffraction limited PSF model. Due to the low values, these may be insecure.}
%\bigskip
\label{tab:analpsf}
\end{minipage}
\end{table*}

\begin{table*}
 \centering
 \begin{minipage}{89mm}
  \caption{Astrometry for SDSS~J1405+0959}
  \begin{tabular}{@{}lrrr@{}}
  \hline 
Object & 
$X [\arcsec]$ &
$Y [\arcsec]$ &
Morphology of GX \\ 
 \hline
A  &  0.000 $\pm$ 0.001 & 0.000 $\pm$ 0.000   & point-like/extended \\
B &  0.248 $\pm$ 0.001 & $-$1.960 $\pm$ 0.001 & point-like/extended \\ 
G1 &  0.048 $\pm$ 0.001  & $-$1.557 $\pm$ 0.001 & point-like/extended \\
G2 &  $-$1.554 $\pm$ 0.009 & $-$0.805 $\pm$ 0.001  & extended  \\
G2 & $-$1.557 $\pm$ 0.008  & $-$0.814 $\pm$ 0.009  & point-like \\
GX & $-$1.789 $\pm$ 0.023 & $-$0.776 $\pm$ 0.013  & extended \\ 
GX & $-$1.812 $\pm$ 0.011  & $-$0.772 $\pm$ 0.010 & point-like \\
GY & $-$1.147 $\pm$ 0.007 & $-$1.588 $\pm$ 0.014  & point-like/extended \\
GYa &  $-$0.825 $\pm$ 0.010 & $-$1.054 $\pm$ 0.009 & point-like/extended \\
GYb &  $-$1.134 $\pm$ 0.017 & $-$1.561 $\pm$ 0.033 & point-like/extended \\
GYc &  $-$1.400 $\pm$ 0.009 & $-$1.933 $\pm$ 0.008 & point-like/extended \\
\hline
\end{tabular}
\\ 
{\footnotesize The positive directions of X and Y are West and North, respectively. The astrometry is computed as the average positions from the $HK'$ bands, where the precision is significantly better than in $J$ band. The errors on the average positions of G2, GX, GY, GYa, GYb, GYc are the sample standard deviation from the $HK'$ bands, and the errors on A, B and G1, which have consistent astrometry between bands, are errors on the mean (assuming Gaussian measurement errors). The astrometry errors do not include a residual geometrical distortion error estimated to be $\sim1.1$ mas (Rusu et al., in prep.), but this has been accounted for when calculating the gravitational lensing parameters in Table 7.}
%\bigskip
\label{tab:1405astrometry}
\end{minipage}
\end{table*}

%%%%%%%%%%%%%%%%%%%%%%%%%%%%%%%%%%%%%%%%%%%%%%%%%%%%%%%%%%%%%%%%%%%%%%

\begin{table*}
 \centering
 \begin{minipage}{101mm}
  \caption{Photometry for SDSS~J1405+0959}
  \begin{tabular}{@{}lcccr@{}}
  \hline 
Object & 
$J$ &
$H$ &
$K'$ &
Morphology of GX \\ 
 \hline
A & 17.98 $\pm$ 0.06 & 17.71 $\pm$ 0.01 & 17.13 $\pm$ 0.04 & point-like/extended \\  %(0.01)
B & 19.06 $\pm$ 0.06 & 18.80 $\pm$ 0.02 & 18.18 $\pm$ 0.04 & point-like/extended \\ %(0.01)
G1 & 17.83 $\pm$ 0.09 & 16.88 $\pm$ 0.06 & 16.22 $\pm$ 0.09 & extended \\
G1 & 17.82 $\pm$ 0.08 & 16.90 $\pm$ 0.06 & 16.27 $\pm$ 0.09 & point-like \\
G2 & 19.14 $\pm$ 0.11 & 18.20 $\pm$ 0.10 & 17.62 $\pm$ 0.09 & extended \\
G2 & 19.00 $\pm$ 0.08 & 18.16 $\pm$ 0.10 & 17.53 $\pm$ 0.08 & point-like \\
GX & 20.74 $\pm$ 0.29 & 20.11 $\pm$ 0.12 & 18.96 $\pm$ 0.12 & extended \\
GX & 21.56 $\pm$ 0.04 & 20.40 $\pm$ 0.05 & 19.39 $\pm$ 0.06 & point-like \\
%GY & 20.29 $\pm$ 0.09 & 19.54 $\pm$ 0.07 & 18.65 $\pm$ 0.07 & GX point-like/extended \\
GY & 20.33 $\pm$ 0.05 & 19.60 $\pm$ 0.04 & 18.75 $\pm$ 0.04 & point-like/extended \\
\hline
\end{tabular}
\\ 
{\footnotesize All photometry is model-fitted photometry estimated with Galfit. However, aperture photometry is given for GY, as described in Section \ref{section:photoz}. Unless written separately, errors on the magnitudes include the scatter between GY being extended/point-like (implemented as Sersic profile and PSF, respectively) or GY consisting of one or three components. The errors on the magnitudes of A and B ($JK'$) signify systematic offset resulting from the simulations, and were not used for estimating the flux ratio error.}
%\bigskip
\label{tab:1405photometry}
\end{minipage}
\end{table*}

%%%%%%%%%%%%%%%%%%%%%%%%%%%%%%%%%%%%%%%%%%%%%%%%%%%%%%%%%%%%%%%%%%%%%%%
 
\begin{figure*}
\includegraphics[width=130mm]{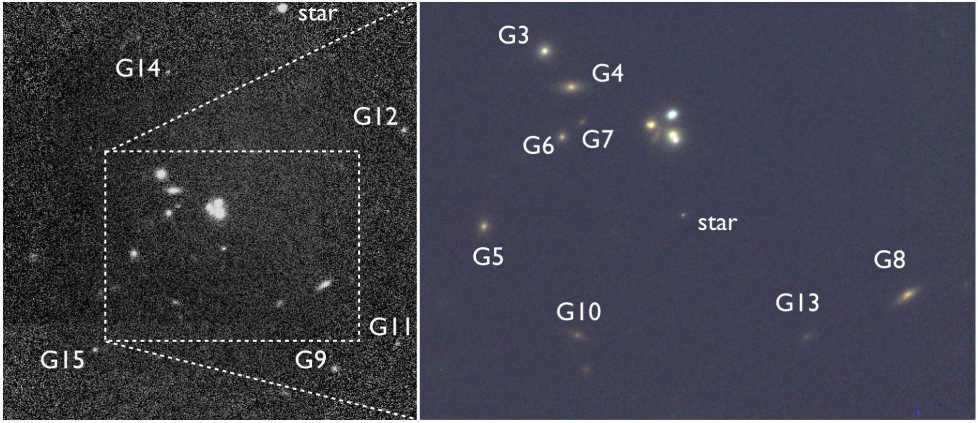}
\caption{Environment of SDSS~J1405+0959 in $J$ band. The field is $70\arcsec\times70\arcsec$ $\mathit{(left)}$. The caption is a $JHK'$ color-combined image. North is up and East is to the left.
\label{fig:1405fov}}
\end{figure*}
%%%%%%%%%%%%%%%%%%%%%%%%%%%%%%%%%%%%%%%%%%%%%%%%%%%%%%%%%%%%%%%%%%%%%%%

%%%%%%%%%%%%%%%%%%%%%%%%%%%%%%%%%%%%%%%%%%%%%%%%%%%%%%%%%%%%%%%%%%%%%%

\begin{table*}
 \centering
 \begin{minipage}{89mm}
  \caption{Morphological parameters for SDSS~J1405+0959}
  \begin{tabular}{@{}ccccc@{}}
  \hline 
Object & 
$e$ &
$PA$ &
$r_e$ [$\arcsec$] &
$n$ \\ 
 \hline
G1 ($J$) & 0.17 $\pm$ 0.03 & 40.3 $\pm$ 3.1 & 0.59 $\pm$ 0.11 & 4.20 $\pm$ 0.65 \\
G1 ($H$) & 0.23 $\pm$ 0.02 & 41.1 $\pm$ 2.7 &1.01 $\pm$ 0.11 & 6.30 $\pm$ 0.33 \\
G1 ($K'$) & 0.18 $\pm$ 0.03 & 43.4 $\pm$ 7.6 & 1.06 $\pm$ 0.30 & 7.49 $\pm$ 0.54 \\
\hline
G2 ($J$) & 0.49 $\pm$ 0.08 & $-$6.7 $\pm$ 3.0 & 0.33 $\pm$ 0.05 & 2.51 $\pm$ 0.55 \\
G2 ($H$) & 0.48 $\pm$ 0.06 & $-$9.0 $\pm$ 1.8 & 0.32 $\pm$ 0.04 & 6.06 $\pm$ 0.50 \\
G2 ($K'$) & 0.54 $\pm$ 0.07 & $-$3.8 $\pm$ 2.8 & 0.25 $\pm$ 0.03 & 5.46 $\pm$ 0.66 \\
\hline
%GX ($J$) & 0.91 $\pm$ 0.14 & 66.1 $\pm$ 7.2 & 0.15 $\pm$ 0.04 & [1]$^a$ \\
%GX ($H$) & 0.66 $\pm$ 0.23 & 32.0 $\pm$ 9.0 & 0.07 $\pm$ 0.02 & 1.07  $\pm$ 3.22  \\
%GX ($K'$) & 0.89 $\pm$ 0.08 & 52.7 $\pm$ 4.8 & 0.11 $\pm$ 0.02 & [1] \\
%\hline
GY ($J$) & 0.80 $\pm$ 0.02 & $-$22.1 $\pm$ 2.1  & 0.64 $\pm$ 0.05 & 0.60 $\pm$ 0.24 \\
GY ($H$) & 0.81 $\pm$ 0.02 & $-$25.2 $\pm$ 1.1  & 0.59 $\pm$ 0.05 & 0.76 $\pm$ 0.17 \\
GY ($K'$) & 0.81 $\pm$ 0.01 & $-$23.8 $\pm$ 1.0  & 0.57 $\pm$ 0.03 & 0.90 $\pm$ 0.13 \\
\hline
\end{tabular}
\\ 
{\footnotesize All four objects are modeled with Sersic profiles. Here $e$ is the ellipticity, $r_e$ is the effective radius measured along the major axis, and $n$ is the Sersic index. The position angle is positive from North towards East. The morphology of G2 is affected little by whether GX is point-like/extended, the results being consistent at $\sim1\sigma$, whereas the other objects are unaffected by this. %$^a$ In order for the models to converge, the Sersic index was fixed at the best-fit value determined in the $H$ band.
}
%\bigskip
\label{tab:1405morphology}
\end{minipage}
\end{table*}

%%%%%%%%%%%%%%%%%%%%%%%%%%%%%%%%%%%%%%%%%%%%%%%%%%%%%%%%%%%%%%%%%%%%%%

The technique summarized above assumes that the adaptive optics PSF does not vary spatially across the components of the system, which is a reasonable assumption given the small maximum separation $\sim2\arcsec$ of these components. The use of the two concentric Moffat profiles as an analytical PSF approximation, instead of, for example, a Gaussian representing a diffraction-limited core and a Moffat \citep{moffat69} for the wing was found to provide a better approximation for the stars observed at low Strehl ratios in the current adaptive optics observation campaign (Rusu et al., in prep.). Additional detail about the technique above is presented in Rusu et al., in prep. 

In order to account for systematics introduced by the technique above, the error bars in Tables \ref{tab:1405astrometry},  \ref{tab:1405photometry} and \ref{tab:1405morphology} show the largest error obtained from several methods: individual fitting statistical error resulting from the Galfit modeling,  scatter from using two different hybrid PSF core radii (10 pixels and 18 pixels), and scatter (where applicable) from GX and GY being fitted as point source/Sersic profile and one/three components, respectively. Finally, the non-simultaneously observed PSF star in $H$ band was used to simulate the observed system 100 times, where each simulation differs in terms of the added noise, taken from a portion of the observed image that is devoid of sources. Each simulated image was then modeled in a way similar to the technique described above. The errors resulting from these simulations are the standard deviations in the 100 values of each best-fitted parameter around the known true value. These therefore include both systematic and statistical errors, and are typically larger than the errors obtained from the other methods. An implicit assumption in this method is that the observed PSF star can be used as a representation of the actual PSFs in the $J$, $H$, and $K'$ bands, albeit solely for the purpose of estimating realistic error bars. This assumption is partly justified by the similarity of the analytical parameters describing the four PSFs in Table \ref{tab:analpsf}, in particular {\it FWHM1}, {\it e1}, {\it FWHM2}, the flux ratio and the Strehl ratio.

As overall results, we have determined that the morphology of G1 and G2 is consistent with extended objects (galaxies). However, due to the low S/N and close proximity to G2, it cannot be reliably distinguished whether GX is extended or point-like. In case it is fit by a Sersic profile, it has a small effective radius and an unreliably determined Sersic index. On the other hand, GY consists of three features. While the central feature is extended, the morphology of the other two cannot be constrained, and for the purpose of extracting the astrometry, they have been fitted as point sources.

%%%%%%%%%%%%%%%%%%%%%%%%%%%%%%%%%%%%%%%%%%%%%%%%%%%%%%%%%%%%%%%%%%%%%%%

 \section{Photometric redshifts}\label{section:photoz}

Using the new high resolution observations, photometric redshifts were computed and a color-color plot was drawn for the objects seen in the field of view that is shown in Figure \ref{fig:1405fov}. The magnitudes of the isolated galaxies were measured with aperture photometry, and those of G1, G2 and GX were obtained from profile fitting. As GY seems to consist of 3 objects, which all appear to be red in Figure \ref{fig:1405closeup}, and are difficult to model analytically due to the low S/N, aperture photometry was performed assuming they all have the same color. For this, all other objects were removed, using the best-fit model obtained with the hybrid PSF, and the residuals were covered with nearby blank sky. Three different blank sky covers were considered, in order to account for possible systematics. An aperture of 30 pixel radius was chosen for the aperture photometry, centered on GYb. The resulting error bars in all bands are smaller than 0.05 mag. 
 
 According to the color-color plot (Figure \ref{fig:1405color}), the galaxies seen in Figure  \ref{fig:1405fov} show various colors, indicative of different redshifts rather than physical association into a group of galaxies. We estimated photometric redshifts from the $JHK'$ bands using the template-fitting methods implemented in the publicly available software HyperZ \citep{bolzonella00}. As templates, we employed the observed mean spectral energy distributions of galaxies from \citet{bruzual93}.

 In particular, G3, G4 and G6 have a redshift probability (G3 is shown in Figure \ref{fig:1405photoz}) consistent with $z\sim 0.3$, which is the estimated redshift of G3 in the SDSS. On the other hand, G1, G2 and G8 have similar colors and redshifts in agreement with $z\sim 0.66$, the spectroscopically estimated redshift of G1. GX has different estimated colors in the extended/point-source cases, and assuming that it is a galaxy, it would be located at $z>1$. More interestingly, GY appears to have a large redshift, consistent with that of the quasar.

\section{Gravitational lens mass modeling}\label{section:lens}

In this section, we test various mass models by exploring  different scenarios given the complexity of this system. The most important constraint for strong lensing models is astrometrical. Table \ref{tab:1405astrometry} provides high-precision astrometry, as a result of the multi-band adaptive optics observations. We also used as constraint the flux ratio $F_A/F_B$, which is very consistent between bands, and also with the optical flux ratio in \citet{inada14}.  An uncertainty of 10\% was set on the observed flux ratio constraint used for the gravitational lens models \citep{yonehara08}. Even in the case of GX being the third quasar image, since its flux is likely to be affected by chromatic processes, we do not use it as a constraint in our model. The singular isothermal profile has extensively been shown to be a good approximation for the mass distribution in lensing galaxies \citep[e.g.,][and references within]{koopmans09,oguri14}. We therefore considered a singular isothermal sphere (SIS), 
%ellipsoid (SIE), 
or the more general power law profile $\rho(r)\propto r^{-\gamma '}$, as our mass models. We performed all gravitational lens modeling with {\it glafic} \citep{oguri10}.

\subsection{Fitting two and three quasar images}\label{section:lens3}

\begin{figure}
\includegraphics[width=84mm]{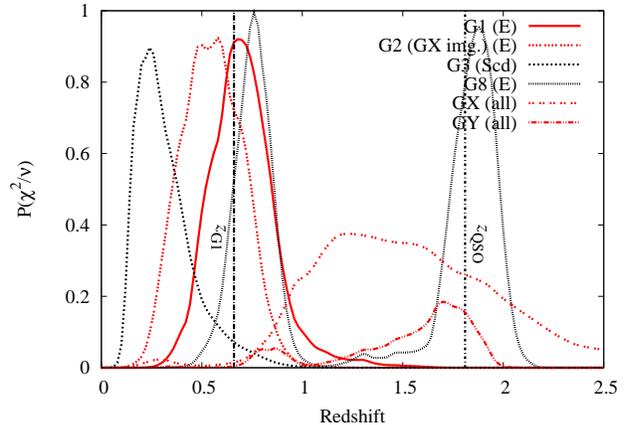}
\caption{Photometric redshifts for galaxies in the SDSS~J1405+0959 FOV, obtained with Hyperz. The labels correspond to those in Figure \ref{fig:1405fov}. The vertical lines mark the spectroscopic redshift of the source quasar and of the main lensing galaxy G1, respectively. The vertical axis represents the probability of exceeding the $\chi^2$ value obtained during the fit by chance, as a function of the number of degrees of freedom (= 2).
\label{fig:1405photoz}}
\end{figure}

Here, we discuss two mass models $a)$ assuming that the system consists of only 2 lensed quasar images A and B, and $b)$ assuming that the system consists of 3 lensed images, A, B, and GX. The latter case is suggested by the insecure morphology of GX (Section \ref{sect:PSF}), as well as the subsequent analysis based on its colors, which we will show in Section \ref{sect:3imag}. First, we considered the case when there are only two quasar images. G2 must be taken into account as a lens, due to its small distance to A and B, which is similar to the separation between them, and its similar luminosity to G1. Given the agreement between the photometric redshift of G1 and G2 (Figure \ref{fig:1405photoz}), we assumed that they are both located at the lens redshift $z_l\sim0.66$, and that the ratio of their velocity dispersions is constrained from their relative luminosities, through the Faber-Jackson relation \citep{faber76}. We used the 2SIS+$\gamma$
model, where the external shear $\gamma$ was considered at the location of G1. We ignored the gravitational lens effect of GX which, assuming it is a galaxy, appears to have different colors and therefore redshift compared to the main deflectors, and is located very close to G2 while being much fainter, making it a comparatively weaker lens. The parameters of this best-fit model, as well as those described below, are given in Table 7. 

Next, the case when GX is an additional quasar image was considered. Here, the observed position of image GX is an additional observational constraint. No flux ratio constraints were used, but the velocity dispersions of both G1 and G2 were fitted as free parameters. A perfect fit (d.o.f. $=0$) was obtained, with the velocity dispersion ratio G2/G1 $=0.76$, %(SIS; for SIE G2/G1 $=0.71$)%
 very close to the value obtained from the Faber-Jackson relation ($\sim0.73$). 
 %In the case of the SIE+$\gamma$ model, the ellipticity of G1 was fixed at a lower value than observed, in order to avoid the production of additional images. 
 This result shows that it is indeed possible for this two-galaxy system to produce three quasar images at the observed positions. 

If an attempt is made to model the observed flux ratio $F_A/F_B$ as well, $\chi^2$/d.o.f. $=4.7/1$ is obtained. While this is not a very good fit, the model might be oversimplified. Allowing for a third order perturbation \citep{bernstein99} a perfect fit is obtained (model 4 in Table 7), where the orientation of the new component is fixed to correspond with that of the shear, in order to avoid a model with negative number of degrees of freedom. The use of a third order perturbation may be justified by the presence of the crowded environment in which the system is embedded (Figure \ref{fig:1405fov}). As a last model, a power law (G1) + SIS (G2) +$\gamma$ profile was also considered, with $\chi^2$/d.o.f.$ =0.3/0$ (model 8 in Table 7). The reason why this is not a perfect fit is that the source is very close to the caustic, which is a curve in the source plane determined from the observed image, lens configuration and lens profile, whose crossing changes the observed image multiplicity (Figure \ref{fig:1405mass}). This can be avoided (i.e. the relative position of the source and caustic can be modified) if ellipticity is also introduced, in the form of the singular isothermal ellipsoid (SIE), at the expense of having negative degrees of freedom: A power law + $\gamma$ (G1) + SIE (G2) model produces $\chi^2$/d.o.f.$ =0/-2$ (model 9 in Table 7). For all mass models described here, elliptical counterparts were also considered (Table 7), where the orientation of the major axis was fixed to the value obtained from morphological fitting (Table \ref{tab:1405morphology}), and the same was done for the ellipticity, unless this cannot reproduce the number of observed images. This approach uses the known correlation between the orientations of mass and light profiles in elliptical lens galaxies \citep[e.g.][]{keeton98,sluse12}, as well as their mass and light ellipticity (\citeauthor{gavazzi12} \citeyear{gavazzi12}; \citeauthor{sluse12} \citeyear{sluse12}, but see also \citeauthor{keeton98} \citeyear{keeton98}; \citeauthor{ferreras08} \citeyear{ferreras08}). Representative lensing models obtained in this Section are plotted in Figure \ref{fig:1405mass}.

%%%%%%%%%%%%%%%%%%%%%%%%%%%%%%%%%%%%%%%%%%%%%%%%%%%%%%%%%%%%%%%%%%%%%%%
\begin{figure}
\includegraphics[width=90mm]{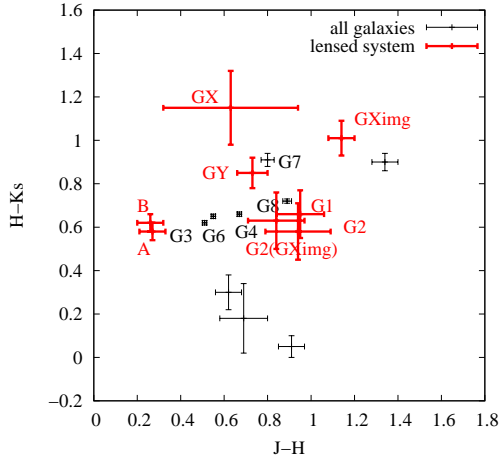}
\caption{Color-color plot of galaxies in the SDSS~J1405+0959 FOV. Objects that are part of the SDSS~J1405+0959 system are marked with red. The labels correspond to those in Figure \ref{fig:1405fov}.
\label{fig:1405color}}
\end{figure}

%%%%%%%%%%%%%%%%%%%%%%%%%%%%%%%%%%%%%%%%%%%%%%%%%%%%%%%%%%%%%%%%%%%%%%%
%%%%%%%%%%%%%%%%%%%%%%%%%%%%%%%%%%%%%%%%%%%%%%%%%%%%%%%%%%%%%%%%%%%%%%%

%%%%%%%%%%%%%%%%%%%%%%%%%%%%%%%%%%%%%%%%%%%%%%%%%%%%%%%%%%%%%%%%%%%%%%%

\subsection{A third quasar image or an additional galaxy?}\label{sect:3imag}
   
As we have shown above, GX is consistent with the position of a third lensed quasar image. Morphological modeling in the $H$ and $K'$ bands cannot distinguish conclusively between a point source and an extended morphology. Furthermore, G2 has an estimated Einstein radius of $\sim0.45\arcsec$ (Table 7), assuming it is located at the redshift of G1, and the separation between G2 and G3 is only $\sim0.26\arcsec$. Since the colors of GX (Figure \ref{fig:1405photoz}) would put it at  $z>1$, in the background of G2, it would be strongly lensed. In fact, four or five images of GX would be produced, some of them brighter and at a larger separation from G2. Such case is clearly inconsistent with the data. 

In view of the above, we discuss the possibility of GX being a quasar image, based on its colors. Although the colors are not consistent with those of images A and B, chromatic differences between quasar images are common in the literature \citep[e.g.,][]{falco99,eliasdottir06,yonehara08}. Such chromatic effects can be caused by dust extinction and microlensing in the lensing galaxy, as well as the intrinsic, time-delayed variability of the source quasar. It is clear that, assuming GX has a point-like morphology, according to Figure \ref{fig:1405color}, it must be reddened in comparison to A and B. Here we test whether this can be explained by dust extinction. We assume that neither image A nor B is affected by extinction, since their colors are very similar. The color difference between A and GX (hereafter renamed image C, when referring to its flux/colors), assumed to be due to extinction, can be used to estimate the extinction, using the equation

%\begin{equation}
\[
m_{C,i}-m_{A,i} = -2.5\log{F_C/F_A}+A_{\lambda,i}\ \ .
\label{eq:surfdens}
\]
%\end{equation}

\noindent Here, the left hand side is the observed magnitude difference of the two images, and $i$ refers each of the three observed bands, $JHK'$. $F$ denotes the model-predicted image flux, and $A_{\lambda}$ is the extinction in the lensing galaxy rest frame, therefore $\lambda=\lambda_{obs}/(1+z_l)$.

To estimate $F_C/F_A$, we use the lensing models from Section \ref{section:lens3}, which predict $F_C/F_A\sim0.25$ (ignoring those with large mass ellipticity). The equation can therefore be solved for $A_{\lambda,i}$, obtaining a value of $\sim2.1$ ($J$), $\sim1.2$ ($H$) and $\sim0.9$ ($K'$). Using an extinction curve from \citet{cardelli89}, and an $\mathrm{R}_\mathrm{V}$ extinction parameter with the Galactic value of $\mathrm{R}_\mathrm{V}=3.1$ (in agreement with the value found from the sample of \'El\'iasd\'ottir), a visual extinction $\mathrm{A}_\mathrm{V}\sim2.9$ ($J$), $\mathrm{A}_\mathrm{V}\sim2.8$ ($H$) and $\mathrm{A}_\mathrm{V}\sim3.1$ ($K'$) is obtained, which translates to $\mathrm{E(B-V)}=\mathrm{A}_\mathrm{V}$/$\mathrm{R}_\mathrm{V}\sim0.9$. Although rare, such large visual extinction and color excess have been observed in lensing galaxies before \citep[e.g.,][]{eliasdottir06,ostman08}. We therefore conclude that color change due to extinction is plausible. A problem with this interpretation however is why image B, located similarly close to the larger G1, also a similar elliptical galaxy, is not similarly affected by extinction.

An alternative explanation is provided by microlensing due to stars in G2, which can also cause large color differences \citep[e.g.,][]{yonehara08}. The small separation between G2 and GX ($\sim1$ effective radius) makes both extinction and microlensing plausible. If GX is the third quasar image, it has a negative parity, and is therefore more sensitive to microlensing effects \citep[e.g.,][]{schechter02}. Microlensing is also expected to produce a flux change that is decreasing with wavelength, as observed. Since neither a spectrum of GX (which would also convincingly show whether it is the third quasar image), nor multi-epoch observations in the same band are available, the relative effects of extinction and microlensing cannot be distinguished using the current data. 

On the other hand, intrinsic variability is not expected to play a significant part, since it generally produces much smaller color differences than observed here \citep[e.g.,][]{yonehara08}, and in addition the similar time delays expected between images B and C (Table 7) would mean that differences should also be seen between A and B. Lastly, we also investigated whether GX is a chance superposition of a nearby star. Such a case is however excluded due to the observed colors, which are inconsistent with any stars observed in the infrared \citep{ducati01}. 

%%%%%%%%%%%%%%%%%%%%%%%%%%%%%%%%%%%%%%%%%%%%%%%%%%%%%%%%%%%%%%%%%%%%%%%

\subsection{More quasar images or a quasar host detection?}\label{sect:lens5}

\begin{figure*}
\includegraphics[width=150mm]{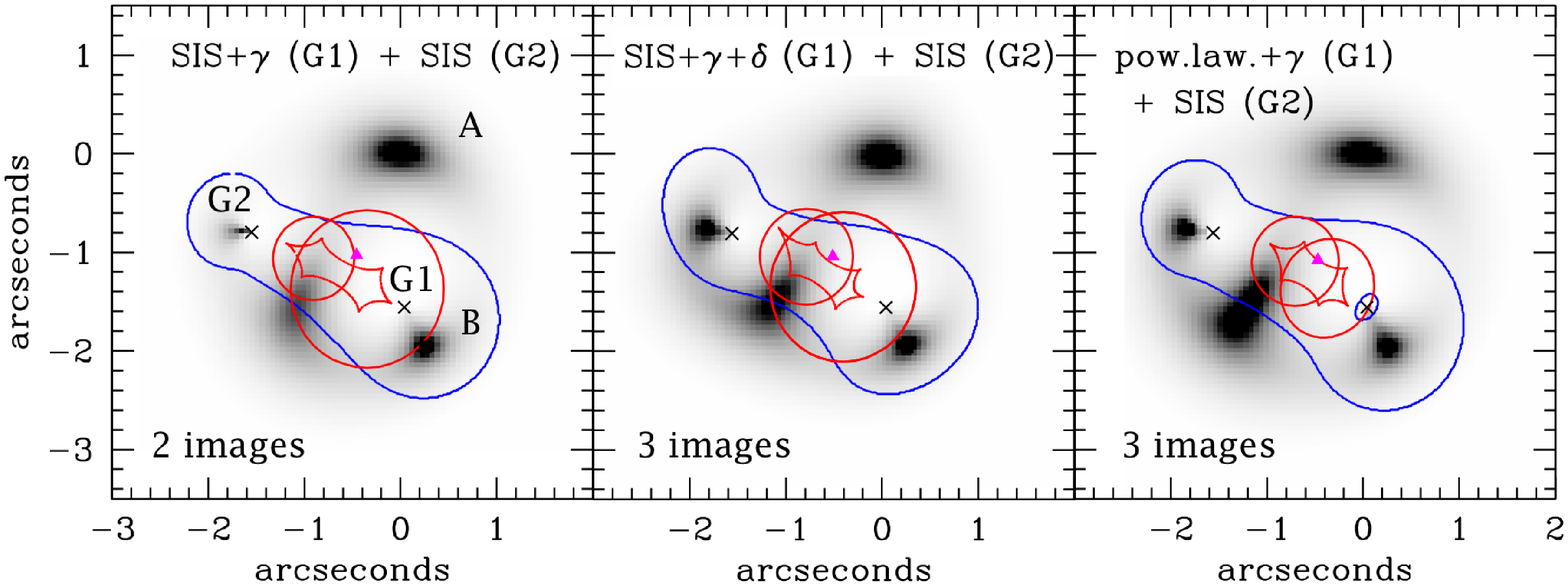}
\caption{ Mass models for SDSS~J1405+0959. The orientation corresponds to that in Figure \ref{fig:1405closeup}. Critical lines $\mathit{(blue)}$ and caustics $\mathit{(red)}$ are shown. The crosses mark the positions of the lensing galaxies, and the triangles the positions of the sources in the source plane. The flux distributions of a lensed circular extended (Sersic profile) source are shown.
\label{fig:1405mass}}
\end{figure*}

From Figure \ref{fig:1405mass}, it is immediately apparent that the source is located very close to a caustic (red curve). As the source crosses inside the caustic, two additional images are created. These images will be located close to, and on each side of the critical line (blue curve). Therefore, it is perhaps not coincidental that the observed elongated galaxy GY is located close to the position where the two additional images are predicted to be. Since morphologically GYb is not consistent with a point source, but only GYa and GYc (although this is not certain, due to the low S/N), the case when GYa and GYc are two additional quasar images was considered. However, this possibility is ruled out because the position of these two objects cannot be reproduced, and in addition they are predicted to be brighter than image A. More specifically, two cases were considered, with or without accounting for GYc as an additional SIS perturber (3SIS+$\gamma$ and 2SIS+$\gamma$, respectively), and only the positions of the 5 images were fitted. This resulted in $\chi^2\sim$1400 for 3 and 4 d.o.f., respectively. 

There is, however, another interpretation for GY. Its photometric redshift suggests that it may be associated with the quasar. As an isolated galaxy at high redshift, it would be lensed by G1 and G2 into multiple images, a fact which is ruled out by the observed data. It could, however, be interpreted as the quasar host galaxy. In fact, as the source is very close to the caustic and the host galaxy is an extended object, a portion of the host galaxy can cross the caustic and be significantly magnified, whereas the point-like quasar does not. Indeed, Figure \ref{fig:1405mass} shows that an extended circular source centered on the source quasar would be lensed into an elongated object whose position, elongation and orientation shows remarkable agreement with GY. 

\begin{figure}
\includegraphics[width=84mm]{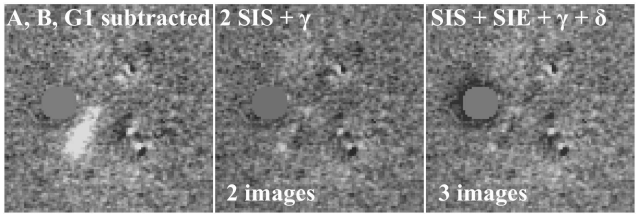}
\caption{ Fitting of GY as a lensed feature in the $K'$ band. {\it Left:} all objects have been fitted and subtracted with {\it glafic}, using an analytical PSF, and without fitting a quasar host galaxy. {\it Center} and {\it right:} a quasar host galaxy modeled as a lensed Sersic profile is also fitted, in the cases when either two or three quasar images are observed, successfully removing most of the flux in GY. Objects G2 and GX, in close proximity to each other, are masked, as the three-image lensing model cannot reproduce the flux of GX.
\label{fig:1405torus}}
\end{figure}

To test this interpretation, we modeled the source as a superposition of a point-like and an extended (elliptical) Sersic profile, and fitted it to the observed data, using {\it glafic}. We used the $K'$ band, where GY is detected at larger S/N, and we modeled the PSF with two concentric Moffat profiles. The reason why the hybrid PSF considered in Section \ref{sect:PSF} is not used is that, through the way it was constructed, it already includes a contribution from the extended source, which cannot be modeled in this case as a different component. We kept fixed the mass model parameters determined from fitting the astrometry and flux ratio constraints in Section \ref{section:lens3}, and varied only the parameters of the PSF and the extended source. Figure \ref{fig:1405torus} depicts the result, which shows that the overall flux of GY is indeed well reproduced by the models, in case of two as well as three observed quasar images. This profile modeling therefore cannot solve the issue of the number of point-like images that are observed, given the current observational data. Faint features are still visible at the peaks of GYa, GYb and GYc, which are not accounted for with this parametric-based source modeling. We computed the total magnification of the extended source models as the flux ratio in the image plane and source plane, and found a model-dependent value of $\sim 6-11$. We also computed the magnification of the modeled GY profile as the ratio of the flux in GY, where all other images were masked, and the region of the source which is mapped into GY (located inside the intersection area of all the caustics shown in Figure \ref{fig:1405mass}). Here we found a magnification factor of $\sim15-20$. The magnification of GY relative to the whole unlensed source is $\sim2-4$.    %(see also Table \ref{tab:lensing}). 

%%%%%%%%%%%%%%%%%%%%%%%%%%%%%%%%%%%%%%%%%%%%%%%%%%%%%%%%%%%%%%%%%%%%%%%
%%%%%%%%%%%%%%%%%%%%%%%%%%%%%%%%%%%%%%%%%%%%%%%%%%%%%%%%%%%%%%%%%%%%%%%

%%%%%%%%%%%%%%%%%%%%%%%%%%%%%%%%%%%%%%%%%%%%%%%%%%%%%%%%%%%%%%%%%%%%%%

\begin{table*}
 \centering
   \rotatebox{90}{
 \begin{minipage}{217mm}
  \caption{Lensing models}
  \begin{tabular}{@{}llllllll@{}}
  \hline 
Model &
Images & 
$\sigma, \theta_E$ & %$\sigma$ [km/s] (M [$h^{-1}  \mathrm{M}_{\astrosun}$]) / $\theta_\mathrm{Ein}$ [$\arcsec$] &
$e, \theta_e$  [deg] &
$\gamma, \theta_\gamma$  [deg] &
$\Delta t$ [days] &
$\mu$ & 
flux ratio \\  
 \hline
  \hline
%\hline \multicolumn{6}{c}{SDSS~J1405+0959} \vspace*{1.5mm} \\ \hline \vspace*{-2.0mm} \\ 
1. SIS$+\gamma$ (G1) & 2 & $252.3\pm^{2.3}_{2.0}, 0.95\pm^{0.02}_{0.01}$  & / & $0.141\pm^{0.007}_{0.006}$, $67.8\pm^{2.5}_{3.6}$ & $129.4\pm^{2.5}_{2.1}$ & $3.20\pm^{0.06}_{0.05}$ & [$0.37\pm^{0.03}_{0.04}$] \\
 \hline
2. 2SIS$+\gamma$ & 2 & $231.9\pm^{1.5}_{2.3}, 0.80\pm^{0.01}_{0.01}$   & / & $0.064\pm^{0.004}_{0.002}$, $54.1\pm^{6.6}_{7.7}$ & $103.6\pm^{1.9}_{1.8}$ & $3.85\pm^{0.10}_{0.10}$ & [$0.37\pm^{0.04}_{0.05}$] \\
& & [$169.3\pm^{1.7}_{1.1}, 0.43\pm^{0.0}_{1.0}$] & & & & & \\
 \hline
3. 2SIE$+\gamma$ & 2 & $233.6\pm^{3.4}_{0.8}, 0.81\pm^{0.01}_{0.01}$   & [$0.18\pm^{0.03}_{0.03}, 43.4\pm^{8.3}_{5.0}$] & $0.053\pm^{0.006}_{0.027}$, $81.2\pm^{5.4}_{11.8}$ & $106.4\pm^{0.8}_{3.1}$ & $3.66\pm^{0.01}_{0.05}$ & [$0.37\pm^{0.04}_{0.01}$] \\
& & [$170.5\pm^{2.5}_{0.6}, 0.43\pm^{0.01}_{0.00}$] & [[0.40], $-3.8\pm^{2.6}_{2.6}$] & & & & \\
%2 img SIE+SIS & $232.1/0.80$  & $0.23^{+0.1}_{-0.1}$, $56.3^{+5.1}_{-7.3}$ & / & $106.8$ & $3.76$ \\
%2 img SIE$+\gamma+$SIS & $230.1, 0.79$   & (0.18, 43.3) & $0.036^{+0.000}_{-0.027}$, $80.7^{+14.3}_{-27.7}$ & $102.2$ (AB) & $3.79$ &  B/A=(0.37) \\
%& (168, 0.42) (G2) & & & & & \\
%2 img + 2 img; SIS$+\gamma$ (G1) + SIE (G2) & $212.8, 0.68$   & / & & $78.5$ (AB) & $4.81$ & B/A=(0.37) \\
%... $\chi^2/\mathrm{d.o.f.}$=0/1 & 208.1, 0.65 (G2) & $0.46$, $94.0$ & $0.064$, $17.0$ & & & \\
 \hline
%3 img (no flux); 2SIS$+\gamma$ & $237.5, 0.84$   & / & $0.092$, $30.0$ & $108.3$(AB) & $5.08$ & B/A=0.25 \\
%& $181.2, 0.49$ (G2) & & & $71.7$ (AC) &  & C/A=0.23 \\
%3 img (no flux) SIE$+\gamma+$SIS & $243.5, 0.89$ & $\equiv 0.11, (43.3)$ & $0.089$, $23.1$ & $114.9$ (AB) & $5.36$ & B/A=0.20 \\
%& $172.2, 0.44$ (G2) & & & $62.7$ (AC) & & C/A=0.25 \\
4. 2SIS$+\gamma+\delta$ & 3 & $226.0\pm^{5.0}_{2.6}, 0.76\pm^{0.03}_{0.02}$ & / & $0.030\pm^{0.016}_{0.009}$, $35.6\pm^{3.1}_{3.9}$ &  $94.1\pm^{5.7}_{3.0}$ & $5.36\pm^{0.21}_{0.46}$ & [$0.37\pm^{0.03}_{0.04}$] \\
 ... ($\delta,\theta_\delta$)=($0.035\pm^{0.009}_{0.024}$, $[35.6\pm^{3.1}_{3.9}]$) & & $181.2\pm^{10.3}_{7.7}, 0.49\pm^{0.07}_{0.04}$ & & & $76.8\pm^{11.1}_{10.0}$ & & $0.28\pm^{0.02}_{0.06}$ \\
 \hline
5. 2SIE$+\gamma+\delta$ & 3 & $234.0\pm^{5.0}_{4.4}, 0.82\pm^{0.03}_{0.05}$ & [$0.18\pm^{0.04}_{0.01}, 43.4\pm^{6.8}_{4.5}$] & $0.017\pm^{0.005}_{0.013}$, $-77.9\pm^{11.9}_{9.5}$ &  $105.4\pm^{7.2}_{9.9}$ & $5.02\pm^{0.11}_{0.29}$ & [$0.37\pm^{0.03}_{0.03}$] \\
 ... ($\delta,\theta_\delta$)=($0.030\pm^{0.009}_{0.007}$, $[-77.9\pm^{14.0}_{8.7}]$) & & $186.0\pm^{9.7}_{7.0}, 0.52\pm^{0.06}_{0.04}$ & [$0.54\pm^{0.04}_{0.09}, -3.8\pm^{3.2}_{1.9}$] & & $81.1\pm^{12.3}_{7.7}$ & & $0.44\pm^{0.08}_{0.17}$ \\
 \hline
6. 2SIS$+\gamma+\delta$; ($\chi^2/\mathrm{d.o.f.}$=0.25/1) & 3+2 & $225.0\pm^{2.0}_{3.8}, 0.76\pm^{0.01}_{0.03}$ & / & $0.032\pm^{0.002}_{0.014}$, $35.6\pm^{7.3}_{6.0}$ &  $92.7\pm^{2.2}_{3.7}$ & $5.35\pm^{0.14}_{0.14}$ & [$0.37\pm^{0.02}_{0.05}$] \\
 ... ($\delta,\theta_\delta$)=($0.036\pm^{0.005}_{0.004}$, $36.8\pm^{1.7}_{1.5}$) & & $182.1\pm^{4.7}_{1.7}, 0.50\pm^{0.02}_{0.02}$ &  & & $77.8\pm^{5.9}_{3.6}$ & & $0.28\pm^{0.02}_{0.04}$ \\
 \hline
7. 2SIE$+\gamma+\delta$; ($\chi^2/\mathrm{d.o.f.}$=0.9/1) & 3+2 & $222.8\pm^{4.6}_{0.5}, 0.74\pm^{0.03}_{0.00}$ & [$0.18\pm^{0.01}_{0.04}, 43.4\pm^{6.8}_{5.0}$] & $0.018\pm^{0.000}_{0.015}$, $-57.2\pm^{36}_{24}$ &  $88.8\pm^{6.0}_{0.0}$ & $5.27\pm^{0.10}_{0.10}$ & [$0.37\pm^{0.03}_{0.02}$] \\
 ... ($\delta,\theta_\delta$)=($0.045\pm^{0.002}_{0.007}$, $43.7\pm^{1.0}_{2.1}$) & & $183.0\pm^{0.9}_{4.9}, 0.50\pm^{0.01}_{0.03}$ & [[0.20], $-3.8\pm^{2.8}_{3.0}$] & & $83.6\pm^{0.7}_{6.8}$ & & $0.32\pm^{0.04}_{0.02}$ \\
 \hline
8. pow. law$ +\gamma$ (G1) + SIS (G2) & 3 &  $0.83\pm^{0.00}_{0.06}$  & / & $0.063\pm^{0.004}_{0.014}$, $31.0\pm^{4.6}_{0.8}$ &  $88.4\pm^{9.2}_{1.2}$ & $6.73\pm^{0.0}_{1.69}$ & [$0.37\pm^{0.02}_{0.05}$] \\
 ... $\gamma'=1.83\pm^{0.14}_{0.00}$; ($\chi^2/\mathrm{d.o.f.}=0.3/0$) & & $174.9\pm^{15.2}_{0.0}, 0.46\pm^{0.09}_{0.00}$ & & & $67.5\pm^{16.2}_{0.0}$ & & $0.21\pm^{0.02}_{0.01}$ \\
 \hline
9. pow. law$ +\gamma$ (G1) + SIE (G2) & 3 &  $0.81\pm^{0.00}_{0.08}$  & $0\pm^{0.09}_{0.00}$, [$43.4\pm^{3.1}_{10.9}$] & $0.063\pm^{0.009}_{0.024}$, $37.6\pm^{19.6}_{2.3}$ &  $90.9\pm^{16.4}_{0.0}$ & $6.99\pm^{0.00}_{1.70}$ & [$0.37\pm^{0.00}_{0.06}$] \\
 ... $\gamma'=1.82\pm^{0.15}_{0.00}$; ($\chi^2/\mathrm{d.o.f.}=0/-2$) & & $169.6\pm^{15.9}_{0.1}, 0.48\pm^{0.02}_{0.06}$ & $0.24\pm^{0.34}_{0.19}$, [$-3.8\pm^{4.0}_{0.6}$] & & $63.8\pm^{14.2}_{2.4}$ & & $0.28\pm^{0.16}_{0.06}$ \\
\hline
\end{tabular}
\\ 
{\footnotesize All external shears $\gamma$ are considered at the position of galaxy G1. For each model, the first (second) line on the third and fourth columns refers to the parameters of G1 (G2); also, the first (second) line on the sixth and eighth columns refers to AB (AC) and $F_B/F_A$ ($F_C/F_A$), respectively. The position angle is positive from North towards East. Values inside square brackets are constrained with a gaussian prior given by the observed uncertainties. Unless otherwise specified, all models have $(\chi^2/\nu=0/0)$. $\sigma$ represents the velocity dispersion of the singular isothermal mass model, $\theta_E$ is the Einstein radius, $e$ is the ellipticity, $\Delta t$ is the time delay and $\mu$ represents the total magnification factor; $\delta$ and $\gamma'$ represent the third order perturbation to the potential, and the logarithmic density slope, respectively. The error bars were determined using the Markov Chain Monte Carlo method implemented in {\it glafic}, where multiple chains were checked for convergence based on the method described in \citet{gelman95}. External convergence from the environment is not considered in the models.}
%\bigskip
\end{minipage}}
\label{tab:lensing}
\end{table*}

%%%%%%%%%%%%%%%%%%%%%%%%%%%%%%%%%%%%%%%%%%%%%%%%%%%%%%%%%%%%%%%%%%%%%%

\begin{figure}
\includegraphics[width=74mm]{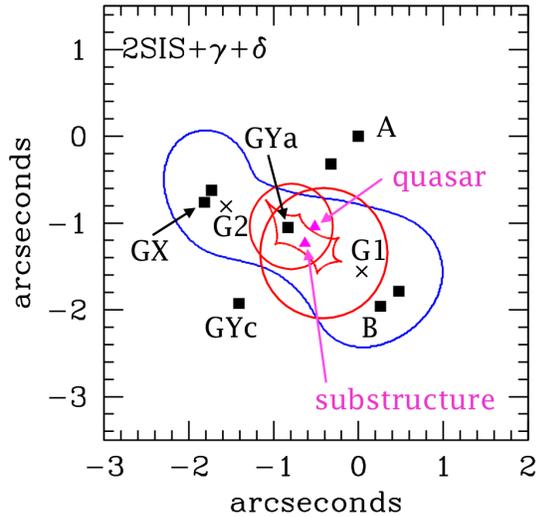}
\caption{ Mass modeling with two sources, simultaneously reproducing A, B, GX as quasar images, and GYa, GYb as part of a multiply lensed (5 image) substructure in the quasar host galaxy.
\label{fig:5imag}}
\end{figure}

The analysis above shows that the data is fully consistent with GY being the quasar host galaxy, lensed into an elongated arc. A question that arises with this interpretation is why GY consists of three components. This might be due to luminous substructure in the host galaxy, irregular overall morphology, a merging system, or small perturbers in the lens plane, which might disturb the arc. In particular, since GYa and GYb are on each side of the critical line, it is plausible that they are mirror-symmetric images of the same feature. This was successfully checked by using one of the mass models predicting three quasar images, and considering a second point-like source to be lensed into GYa and GYb (Figure \ref{fig:5imag}). The use of a second source sets tighter constraints on the mass model (models 6 and 7 in Table 7). Three additional images are predicted, and their non-detection can presumably be explained by their close proximity to the quasar images, and their expected small fluxes. The two sources are separated by $\sim0\farcs22$ in the source plane, or $\sim2$ kpc. This is on the order of one effective radius for typical galaxies at the source redshift \citep[e.g.,][]{trujillo07}. We also tried to fit two sources to the mass models predicting two quasar images, but this produced poor fits, with large lens ellipticity for G1 and G2, at vastly different orientations than the observed light profiles.

In conclusion, in view of the photometric redshifts, as well as the good overall fit with the flux distribution expected for an extended source crossing a caustic, the data implies that GY indeed represents part of the quasar host galaxy. We refrain however from making quantitative estimates of the physical properties of the host galaxy from the current data. This is because of the quasar image number (and therefore the magnification factor) being unconfirmed, the fragmentation of GY, not accurately reproduced by a single analytical source profile, and the fact that it is difficult to assess whether the analytical PSF is reliably disentangled from, or biases the parameters of the extended source.

%%%%%%%%%%%%%%%%%%%%%%%%%%%%%%%%%%%%%%%%%%%%%%%%%%%%%%%%%%%%%%%%%%%%%%%
%%%%%%%%%%%%%%%%%%%%%%%%%%%%%%%%%%%%%%%%%%%%%%%%%%%%%%%%%%%%%%%%%%%%%%%

%%%%%%%%%%%%%%%%%%%%%%%%%%%%%%%%%%%%%%%%%%%%%%%%%%%%%%%%%%%%%%%%%%%%%%%
%%%%%%%%%%%%%%%%%%%%%%%%%%%%%%%%%%%%%%%%%%%%%%%%%%%%%%%%%%%%%%%%%%%%%%%

%\section{Comparison with MG~2016+112}\label{sect:More}

%\begin{landscape}
%\begin{figure}
%\rotate
%\epsscale{1}
%\includegraphics[width=84mm]{environ2.eps}
%\caption{SDSS environment and redshift cuts for selected systems. See the text for the redshift values. All galaxies kept after the cut have $i<21$ and photometric redshift error $\Delta z < 0.2$, ensuring accurate photometric redshifts \citep{csabai03}. The cuts satisfy at least one of the two redshift estimates in the SDSS database, PHOTOZ1 and PHOTOZ2.
%\label{fig:environ1}}
%\end{figure}
%\end{landscape}
%%

\section{Discussion, conclusions and future works}\label{sect:concl}

We have conducted high resolution, near-infrared, multi-band adaptive optics observations with the Subaru Telescope of the gravitationally lensed quasar SDSS~J1405+0959. Despite the low Strehl ratios, the observations were successful in uncovering unprecedented detail in this system, and in particular in discovering two new components named GX and GY. We used a new technique to derive the PSF, and determine the morphology, astrometry and photometry of the individual components that constitute the observed system.

We showed that GX is likely to be the third quasar image in this system, based on its colors as well as lens modeling. If confirmed, it will be the first small-separation three-image lensed quasar produced by a pair of lensing galaxies, and in addition will allow the study of the interstellar medium of G2. The nature of GX may be tested via high spatial resolution spectroscopy, or alternatively via deep observations with the Hubble Space Telescope ($HST$), which produces a better characterized PSF, and may distinguish between the point-source or extended morphology of GX. In addition, such observations would also better separate the extended quasar host emission from the point-like quasar source. We note that our technique to obtain the PSF may be affected by faint host emission underneath the bright quasar image. This might be responsible for some of the discrepancies in the morphology of G1 and G2 in the three observed bands, since the host galaxy is more prominent at longer wavelengths.     

 We also showed that GY consists of additional lensed images of an extended region of the quasar host galaxy, based on the observed colors and the match between the observed light distribution and that expected for an extended lensed source, given the best available lensing models. Adaptive optics-assisted or $HST$ spectroscopy to determine a redshift would unambiguously confirm the nature of GY, as well as allow us to study its physical properties, such as star formation. The added advantage here is that GY shows emission from the quasar host galaxy alone which is free of contamination from the bright quasar components. Deeper adaptive optics observations, to increase the S/N of GY, would provide hundreds of intensity pixels to further study the mass distribution \citep[e.g.,][]{suyu09} in this two-galaxy lens system. This would allow a non-parametric reconstruction of the quasar host galaxy in the source plane, which would be helpful to explain the fragmentation observed in GY and, for example, to study the correlation with the supermassive black hole \citep[e.g.,][]{peng06}. 
 
As an analogy, the known system most similar to SDSS~J1405+0959 is MG 2016+112, which also contains an observed arc without a central quasar image. That system also requires two lens galaxies to model, although the second lens is comparatively much weaker, with Einstein radius ratio G1/D $\sim0.11$ (following the notations in the literature), whereas for SDSS~J1405+0959, G2/G1 $\sim0.55$. As such, the tangential caustics in SDSS~J1405+0959 merge and the two radial caustics overlap, whereas this does not occur in the case of MG 2016+112, which only produces two quasar images. The latter system however contains much more astrometric constraints, as radio observations \citep{more09} reveal mirror-symmetric sub-images in both quasar images, as well as in the extended arc. On the other hand, SDSS~J1405+0959 is radio-quiet, based on non-detection in existing radio surveys. 

%%%%%%%%%%%%%%%%%%%%%%%%%%%%%%%%%%%%%%%%%%%%%%%%%%%%%%%%%%%%%%%%%%%%%%%

\section*{Acknowledgments}

This work was supported in part by the Japan Society for the Promotion of Science (JSPS) KAKENHI Grant Number 26800093 and World Premier International Research Center Initiative (WPI Initiative), Ministry of Education, Culture, Sports, Science and Technology (MEXT), Japan. C.E.R. acknowledges the support of the JSPS Research Fellowship.

The authors would like to thank Sherry Suyu for helpful discussions.
 %as well as the anonymous referee for criticism that helped improve the paper. 
 The authors recognize and acknowledge the very significant cultural role and reverence that the summit of Mauna Kea has always had within the indigenous Hawaiian community. We are most fortunate to have the opportunity to conduct observations from this superb mountain.

%%%%%%%%%%%%%%%%%%%%%%%%%%%%%%%%%%%%%%%%%%%%%%%%%%%%%%%%%%%%%%%%%%%%%%%

%\appendix

%\section[]{Large gaps in L\lowercase{y}${\balpha}$ forests\\* due to fluctuations in line distribution}
%\begin{equation}
%x={(1+z)^{\gamma+1}-(1+z_1)^{\gamma+1} \over
%     (1+z_2)^{\gamma+1}-(1+z_1)^{\gamma+1}}.
%\end{equation}
%\newpage
%\begin{figure}
%\vspace{11pc}
%\caption{$P(>x_{\rmn{gap}})$ as a function of $x_{\rmn{gap}}$ for,
 %from left to right, $N=160$, 150, 140, 110, 100, 90, 50, 45 and~40.
% Compare this with \protect\citet{b15}.}
%\label{appenfig}
%\end{figure}

%frequency $\nu$ 
%\begin{equation}
%   L(\nu)=\mskip-12mu\int\limits_{\rmn{envelope}}\mskip-12mu
%   \rho(r)Q_{\rmn{abs}}(\nu)B[\nu,T_{\rmn{g}}(r)]\exp [-\tau(\nu,r)]\>
%   \rmn{d}V,
%\end{equation}
% where
% $Q_{\rmn{abs}}(\nu)$ is the absorption efficiency at frequency $\nu$,
% $\rho(r)$            is the dust grain density,
% $T_{\rmn{g}}(\nu)$    is the grain temperature,
% $B[\nu,T_{\rmn{g}}(r)]$  is the Planck function, and
% $\tau(\nu,r)$        is the optical depth at distance {\it r\/} from the
%                      centre of the star.

%\[
%  [\nu_1]-[\nu_2]=-2.5\log [f(\nu_1)/f(\nu_2)],
%\]
%[12]--[25] 

\bsp

\label{lastpage}

\begin{thebibliography}{99}

\bibitem[Abazajian et al.(2009)]{abazajian09} %
Abazajian,K.~N., Adelman-McCarthy, J.~K., Ag\"{u}eros, M. A., et al. 2009, ApJS, 182, 543

\bibitem[Bernstein et al.(1999)]{bernstein99} %
Bernstein, G., Fischer, P.\ 1909, AJ, 118, 14 

\bibitem[Bolzonella et al.(2000)]{bolzonella00} %
Bolzonella, M., Miralles, J.-M., \& Pell{\'o}, R.\ 2000, A\&A, 363, 476 

\bibitem[Bruzual \& Charlot(1993)]{bruzual93} %
Bruzual, A., G., \& Charlot, S.\ 1993, ApJ, 405, 538 

\bibitem[Cardelli et al.(1989)]{cardelli89} %
Cardelli, J.~A. Clayton, G.~C., Mathis, J.~S.\ 1989, ApJ, 345, 245 

\bibitem[Ducati et al.(2001)]{ducati01} %
Ducati, J.~R. et al.\ 2001, ApJ, 558, 309

\bibitem[\'El\'iasd\'ottir et al.(2006)]{eliasdottir06} %
\'El\'iasd\'ottir, A. et al.\ 2006, ApJS, 166, 443

\bibitem[Faber \& Jackson(1976)]{faber76} %
Faber, S.~M., \& Jackson, R.~E.\ 1976, ApJ, 204, 668

\bibitem[Falco et al.(1999)]{falco99} %
Falco, E.~E.\ 1999, ApJ, 523, 617

\bibitem[Ferreras et al.(2008)]{ferreras08} %
Ferreras, I., Saha, P., \& Burles, S.\ 2008, MNRAS, 383, 857

\bibitem[Gavazzi et al.(2012)]{gavazzi12} %
Gavazzi, R. et al.\ 2012, ApJ, 761, 170

\bibitem[Gelman et al.(1995)]{gelman95} %
Gelman, A. et al.\ 1995, Bayesian Data Analysis, New York: Chapman \& Hal/CRC

\bibitem[Hayano et al.(2008)]{hayano08} %
Hayano, Y., et al.\ 2008, Proc.~SPIE, 7015, 25

\bibitem[Hayano et al.(2010)]{hayano10} %
Hayano, Y., et al.\ 2010, Proc.~SPIE, 7736, 21

\bibitem[Inada et al.(2012)]{inada12} %
Inada, N. et al., 2012, AJ, 143, 119

\bibitem[Inada et al.(2014)]{inada14} %
Inada, N., et al.\ 2014, AJ, 147, 153

\bibitem[Jackson et al.(2012)]{jackson12} %
Jackson, N. et al., 2012, MNRAS, 419, 2014

\bibitem[Keeton et al.(1998)]{keeton98} %
Keeton, C.~R., Kochanek, C.~S., \& Falco, E.~E.,\ 1998, ApJ, 509, 561

\bibitem[Kobayashi et al.(2000)]{kobayashi00} %
Kobayashi, N., et al. \ 2000, Proc.~SPIE, 4008, 1056

\bibitem[Koptelova et al.(2013)]{koptelova13} %
Koptelova, E. et al,\ 2013, Preprint: arXiv:astro-ph/1307.2390

\bibitem[Koopmans et al.(2002)]{koopmans02} %
Koopmans, L.,~V.E. et al,\ 2002, MNRAS, 334, 39

\bibitem[Koopmans et al.(2009)]{koopmans09} %
Koopmans, L.,~V.E. et al,\ 2009, ApJ, 703, L51

\bibitem[Kuhlbrodt et al.(2005)]{kuhlbrodt05} %
Kuhlbrodt, B. et al,\ 2005, A\&A, 439, 497

\bibitem[Lagattuta et al. (2010)]{lagattuta10} %
Lagattuta, D.~J., Auger, M.~W., Fassnacht, C.~D. \ 2010, ApJ, 716, L185

\bibitem[Lawrence et al.(1984)]{lawrence84} %
Lawrence, C.,~R., et al,\ 1984, Science, 223, 4631, 46

\bibitem[Lewis et al.(2002)]{lewis02} %
Lewis, G.,~F., et al,\ 2002, MNRAS, 334, L7

\bibitem[Minowa et al.(2005)]{minowa05} %
Minowa, Y., et al.\ 2005, ApJ, 629, 29 

%\bibitem[Minowa et al.(2010)]{minowa10} %
%Minowa, Y., et al.\ 2010, Proc.~SPIE, 7736, 122

\bibitem[Minowa et al.(2012)]{minowa12} %
Minowa, Y., et al.\ 2012, Proc.~SPIE, 8447, 1

\bibitem[Moffat(1969)]{moffat69} %
Moffat, A.~F.~J.,\ 1969, A\&A, 3, 455 

\bibitem[More et al.(2009)]{more09} %
More, A. et al. \ 2009, MNRAS, 394, 174

\bibitem[Oguri et al. (2006)]{oguri06} %
Oguri, M., et al. \ 2006, AJ, 132, 999

\bibitem[Oguri et al. (2008)]{oguri08} %
Oguri, M., et al. \ 2008, ApJ, 676, L1

\bibitem[Oguri (2010)]{oguri10} %
Oguri, M.\ 2010, PASJ, 62, 1017

\bibitem[Oguri et al.(2014)]{oguri14} %
Oguri, M., Rusu, C.~E., Falco, E.~E.\ 2014, MNRAS, 439, 2494

\bibitem[Ostman et al.(2008)]{ostman08} %
Ostman, L., Goobar, A., Mortsell, E.\ 2008, A\&A, 485, 403

\bibitem[Oya et al.(2013)]{oya13} %
Oya, S., et al. \ 2013, PASJ, 65, 1, 9

\bibitem[Peng et al.(2002)]{peng02} %
Peng, C.~Y., Ho, L.~C., Impey, C.~D., \& Rix, H.-W.\ 2002, AJ, 124, 266 

\bibitem[Peng et al.(2006)]{peng06} %
Peng, C.~Y. et al. 2006, ApJ, 649, 616 

\bibitem[Rusu et al.(2011)]{rusu11} %
Rusu, C.~E. et al. 2011, ApJ, 738, 30 

\bibitem[Rusin et al.(2001)]{rusin01} %
Rusu, C.~E. et al. 2001, ApJ, 557, 594 

\bibitem[Schechter et al.(2002)]{schechter02} %
Schechter, P.~L. \& Wambsganss, J. \ 2002, ApJ, 580, 685

\bibitem[Schlegel et al.(1998)]{schlegel98} %
Schlegel, D.~J., Finkbeiner, D.~P., Davis, M.\ 1998, ApJ, 500, 525 

\bibitem[Shin et al.(2009)]{shin08} %
Shin, E.~M., Evans, N.~W. \ 2008, MNRAS, 390, 505

\bibitem[Sluse et al.(2012)]{sluse12} %
Sluse, D. et al.\ 2012, A\&A, 544, 62 

\bibitem[Suyu et al.(2009)]{suyu09} %
Suyu, S.~H., et al.\ 2009, ApJ, 691, 277 

\bibitem[Trujillo et al.(2007)]{trujillo07} %
Trujillo, I., et al.\ 2007, MNRAS, 382, 109 

\bibitem[Watanabe et al.(2004)]{watanabe04} %
Watanabe, N., et al. \ 2004, Proc.~SPIE, 5490, 1096

\bibitem[Winn et al.(2004)]{winn04} %
Winn, J.,~N., et al,\ 2004, Nat., 334, 427, 613

\bibitem[Yonehara et al.(2008)]{yonehara08} %
Yonehara, A., Hirashita, H., Richter, P.\ 2008, A\&Ap, 478, 95

\bibitem[Zacharias et al.(2004)]{zacharias04} %
Zacharias, N., et al., Bulletin of the American Astronomical Society, Vol. 36, p.1418

%\bibitem[\protect\citeauthoryear{Baird}{1981}]{b1} Baird S.R., 1981, ApJ, 245, 208
%\bibitem[\protect\citeauthoryear{Beichman et al.}{1985b}]{b3} Beichman C.A., Neugebauer G., Habing H.J., Clegg P.E., Chester T.J., 1985b,
%{\it IRAS\/} Explanatory Supplement. Jet Propulsion Laboratory, Pasadena
\end{thebibliography}
\end{document}